\documentclass[pra,twocolumn,floatfix,showpacs,nofootinbib]{revtex4}
\usepackage{rotating,enumerate,longtable,dcolumn,graphicx,graphics,times,amsmath,amsfonts,amssymb,latexsym,bm,bbm,textcomp,dsfont}
\usepackage[T1]{fontenc}
\usepackage[latin9]{inputenc}
\usepackage{amsbsy}

\newcommand{\bra}[1]{\langle#1 |}
\newcommand{\ket}[1]{|#1 \rangle}
\newcommand{\braket}[2]{\left \langle #1 \mid #2 \right \rangle}
\newcommand{\ketbra}[2]{\vert #1 \rangle \! \langle #2 \vert}

\newcommand{\average}[1]{\langle #1 \rangle}

\newcommand{\be}{\begin{equation}}
\newcommand{\ee}{\end{equation}}
\newcommand{\bae}{\begin{eqnarray}}
\newcommand{\eae}{\end{eqnarray}}

\begin{document}
\global\long\def\braket#1#2{\langle#1|#2 \rangle}
 \global\long\def\bra#1{\langle#1|}
 \global\long\def\ket#1{|#1\rangle}
 \global\long\def\expect#1{\langle#1 \rangle}

\title{The transition to chaos of coupled oscillators: An operator fidelity susceptibility study}

\author{N. Tobias Jacobson\footnote{Corresponding author: ntj [at] usc.edu}$^{1}$, Paolo Giorda$^{2}$, and Paolo Zanardi$^{1,2}$}
\affiliation{$^{1}$Department of Physics and Astronomy, University of Southern California, Los Angeles, CA 90089,\\
$^{2}$Institute for Scientific Interchange, Viale Settimio Severo 65, I-10133 Torino, Italy}

\begin{abstract}
The operator fidelity is a measure of the information-theoretic distinguishability between perturbed and unperturbed evolutions. The response of this measure to the perturbation may be formulated in terms of the operator fidelity susceptibility (OFS), a quantity which has been used to investigate the parameter spaces of quantum systems in order to discriminate their regular and chaotic regimes. In this work we numerically study the OFS for a pair of non-linearly coupled two-dimensional harmonic oscillators, a model which is equivalent to that of a hydrogen atom in a uniform external magnetic field. We show how the two terms of the OFS, being linked to the main properties that differentiate regular from chaotic behavior, allow for the detection of this model's transition between the two regimes. In addition, we find that the parameter interval where perturbation theory applies is delimited from above by a local minimum of one of the analyzed terms.
\end{abstract}

\maketitle
\section{Introduction}

Given a classical system in which chaotic dynamics arise, a fundamental
problem is to understand the behavior of the corresponding quantum analogue
of this system. A starting point for this comparison is the observation that while classical dynamical chaos can typically be described in terms of the divergence of initially neighboring trajectories in phase space, the unitary evolution of closed quantum systems does not allow such a characterization. So, how does classical chaos manifest in a quantum system? Is quantum chaos reflected in the distribution of the energy levels of the corresponding quantum system, for example, or in terms of the temporal evolution of suitable expectation values?
For the last several decades, intense effort has been devoted to the study of quantum chaos \cite{Chaos}. A variety of approaches have been taken, including random matrix theory \cite{RandomMtx}, quantum motion reversibility \cite{QCh-rev}, stability \cite{QCh-stab}, fidelity \cite{Prosen}, entanglement \cite{chaosent}, and recently measures  of phase-space growth rates \cite{casati08}. All of these techniques successfully address specific aspects of quantum chaotic behavior.
The fidelity, a quantity commonly encountered in the context of quantum information theory, has been used extensively in the form of the 
Loschmidt echo to study quantum chaos by Prosen, et. al \cite{Prosen} and others. See Ref. \cite{ProsenLoshmidtDynamics} for an extensive review of the 
fidelity approach to quantum chaos. The response of this quantity 
to infinitesimal perturbations, what we call the \emph{operator fidelity susceptibility} (OFS), was formulated from a differential-geometric 
perspective in \cite{WSW,LSWZ}. This quantity has been studied in the context of quantum chaos in \cite{GZ}, where the state- and level-dependent 
contributions to the OFS were distinguished.
The operator fidelity and OFS are the generalizations of the {\em ground state} fidelity and fidelity susceptibility, resepectively, that have been fruitfully used 
\cite{generalstatefid,Zhou_Barjaktarevic,YouLiGu} over the last years to characterize another important class of phenomena encountered
in many-body quantum physics: quantum phase transitions. The state fidelity approach is based on the idea that one can detect quantum critical points by employing a geometric measure of distinguishability between ground states corresponding to neighboring parameters.
The OFS is a natural extension of this approach from the space of states to operators. In an analogous fashion to the ground state case, the OFS gives a measure of the rate of distinguishability between ``neighboring'' members of a family
of operators. Here we consider the unitary evolution operators $U_{\lambda}=e^{-iH(\lambda)t}$
generated by a family of time-independent Hamiltonians $H(\lambda)=H_0+\lambda V$.
With this choice, the OFS measures the rate of separation between
the unitary evolutions induced by $U_{\lambda}$ and $U_{\lambda+\delta\lambda}$, where $\delta\lambda V$ is
an infinitesimally small perturbation to the Hamiltonian.

The reason why the OFS $\chi(\lambda)$
allows one to distinguish the transition from a regular to a chaotic regime of a quantum system is two-fold.
First of all, the theoretical study of $\chi(\lambda)$ carried out in \cite{GZ} and based on random matrix theory arguments \cite{RandomMtx} has shown on general grounds that for systems characterized by random perturbations $V$ drawn from proper ensembles, the quantum chaotic evolutions can be characterized as those which have the highest resilience to these perturbations. 
This can be seen by expressing the OFS in terms of an autocorrelation function of the perturbation \cite{ProsenLoshmidtDynamics}. In this form, 
when the system is chaotic the correlation function may decay more rapidly than in the regular regime, leading to a slower decay of the fidelity.
Furthermore, as described in  \cite{LSWZ,GZ}, and as will be reviewed in the next
section, the OFS can be split into two terms, $\chi(\lambda)=\chi^{(1)}(\lambda)+\chi^{(2)}(\lambda)$, one which depends on the variation of the energy levels and the other on the variation of the eigenvectors.
In particular, the first term depends on the variation of the eigenvector and is directly related to the spacings between energy levels. 
In this respect, $\chi^{(1)}(\lambda)$  is analogous to the ground-state fidelity susceptibility \cite{generalstatefid, YouLiGu}, a quantity which depends on the energy gap between the ground and first excited states. This property links the OFS to one of the main paradigms for the definition of quantum chaos: the statistics of the spacings $s_n=E_{n+1}-E_{n}$ between neighboring energy levels. Indeed, in a seminal paper Bohigas, Giannoni, and Schmidt \cite{BGS} proposed that certain universal level spacing statistics encountered
in random matrix theory \cite{RandomMtx} also arise in the spectra of regular or chaotic quantum systems. In particular, regular systems are expected to have Poissonian level spacing statistics, $P(s)=e^{-s}$.
In this case the most probable value is $s=0$, since frequent crossings between members of uncorrelated subsets of energy levels may occur. This property is a direct consequence of the integrability of the system, corresponding to the existence of a sufficiently large number of conserved quantities. 
On the other hand, with the exact form of the distribution depending on the few existing symmetries of the model, chaotic systems may display Wigner-type statistics, $P(s)\approx s^\beta \exp{-s^2}$. In this case level crossings are suppressed as a result of correlations between the energy levels due to a lack of symmetry in the system.
The sensitivity of the eigenvector part of the OFS to these level statistics has already been successfully tested in \cite{GZ}, where it described the transition to chaos in the Dicke model.
The same basic ideas have subsequently been used to define a simplified version of the OFS that has been applied to spot the avoided crossings in some Bose-Hubbard systems \cite{Plotz}.

In this paper we employ the operator fidelity susceptibility to investigate a system of two non-linearly coupled two-dimensional harmonic oscillators. The importance of this model arises from its equivalence to a prototypical system for which the regimes of classical and quantum chaos have been well-documented theoretically, and for which experimental data exist and agree exceptionally well with theoretical predictions: the hydrogen (or hydrogen-like) atom in a time-independent and uniform external magnetic field \cite{LesHouches,DG,DGJPhysB}.
In order to spot the transition from regularity to chaos, we compute for the oscillator problem both parts of $\chi(\lambda)$. Our analysis, besides providing a further test of the applicability of the OFS, will allow us to compare their different behavior in the transition and to give an interpretation of the full OFS.

In Sec. \ref{sec:Operator-fidelity} we briefly review the operator
fidelity and operator fidelity susceptibility. In Sec. \ref{sec:Model}
we explain how the model is derived, describe the algebraic representation allowing for its efficient diagonalization, and detail the numerical procedure adopted. In Sec. \ref{sec:Results} we outline the
level spacing statistics for the model and give the results of the operator fidelity susceptibility analysis.
We conclude with Sec. \ref{sec:Conclusion}.

\section{\label{sec:Operator-fidelity}Operator fidelity}

Due to the preservation of inner products under unitary evolution,
the classically useful notion of the divergence of trajectories in
phase space resulting from sensitivity to initial conditions does
not apply for quantum systems. One may instead compare the unitary
evolution \emph{operators} corresponding to nearby points in parameter
space. This approach allows the problem of discriminating between the regular and chaotic regimes of a given quantum system to be cast into the information-theoretic idea of statistical distinguishability between states representing evolution operators. 

We start our discussion by recalling that the overlap between quantum states belonging to a $D$-dimensional Hilbert space $\mathcal{H}$ generalizes to the space $\mathcal{L}(\mathcal{H})$ of linear operators acting on $\mathcal{H}$ in a simple way.
Supposing, for example, that $\rho=\sum_i p_i \ketbra{i}{i}$ is the density matrix of a
given quantum system, one can identify any operator $X \in \mathcal{H}$
with the state $\ket{X}\in \mathcal{H}\otimes \mathcal{H}$ defined in terms of the purification of $\rho$,
$\ket{\psi}_\rho=\sum_i \sqrt{p_i} \ket{i}\otimes\ket{i}\in \mathcal{H}\otimes \mathcal{H}$, as
\be
\ket{A}\doteq A\otimes \mathbb{I} \ket{\psi}_\rho.
\ee
The $\rho$-fidelity between two operators $X,Y$ is then the (bi-partite) state fidelity:  \be
{\cal F}_\rho(X,Y)\doteq |\braket{X}{Y}_\rho|=|\mbox{Tr}(\rho X^\dagger Y)|.
\ee
where $\braket{X}{Y}_\rho$ defines an hermitean scalar product over $\mathcal{L}(\mathcal{H})$.
In this work, as in \cite{GZ,LSWZ}, our starting point is to compute
the $\rho$-fidelity between members of the one-parameter family of
unitary time-evolution operators, $U_{\lambda}(t)=e^{-itH_{\lambda}}$,
corresponding to slightly different Hamiltonian parameters $\lambda,\lambda+\delta\lambda$.
The second-order expansion $\ket{U_{\lambda+\delta\lambda}}_\rho= \ket{U_{\lambda}}_\rho + \delta\lambda \ket{\partial_\lambda U_{\lambda}}_\rho + \delta\lambda^2 \ket{\partial^2_\lambda U_{\lambda}}_\rho/2$ of the state representing the perturbed evolution, together with the relations obtainable by the identity $\partial^2_\lambda\braket{U_{\lambda}}{U_{\lambda}}_\rho=0$, allow the second-order expansion of the fidelity to be written
\begin{equation}
\mathcal{F}_{\rho}(U_{\lambda},U_{\lambda+\delta\lambda})=
1-\frac{\delta\lambda^{2}}{2}\chi_{\rho}(\lambda),
\label{eq:OFS}
\end{equation}
in terms of the \emph{operator fidelity susceptibility} (OFS) \cite{LSWZ},
\begin{equation}
\chi_{\rho}(\lambda):=\braket{\partial_{\lambda}U}{\partial_{\lambda}U}_\rho-
\vert\braket{\partial_{\lambda} U_\lambda}{U_\lambda}_\rho\vert^{2}.
\end{equation}
In this work we always choose states $\rho$ which commute with the
Hamiltonian, $H_{\lambda} = H_0 + \lambda V$. The spectral decomposition of $\rho$ in a basis of energy eigenstates of $H_{\lambda}$ and the application of time-dependent perturbation theory \cite{LSWZ} allows the OFS to be decomposed into two parts, $\chi_{\rho}(\lambda)=\chi_{\rho}^{(1)}(\lambda)+\chi_{\rho}^{(2)}(\lambda)$, where \cite{NoteOnNotation}
\begin{eqnarray}
\chi_{\rho}^{(1)}(\lambda) & =&2\pi t \sum_{\genfrac{}{}{0pt}{}{n,m=1}{m \neq n}}^{D}\rho_{n}|\bra n\partial_{\lambda}H\ket m|^{2}G_{t}(E_{n}-E_{m})\label{eq:Chi1}\\
\chi_{\rho}^{(2)}(\lambda) & = & t^{2}\big(\sum_{n=1}^{D}\rho_{n}\vert\bra n\partial_{\lambda}H\ket n\vert^{2}-\nonumber \\
&-&\vert\sum_{n=1}^{D}\rho_{n}\bra n\partial_{\lambda}H\ket n\vert^{2}\big),\label{eq:Chi2}\end{eqnarray}
$D$ is the dimension of the Hilbert space, and
$G_{t}(x)\equiv\frac{2\sin^{2}(tx/2)}{\pi tx^{2}}$.
Following the discussion in Ref. \cite{LSWZ}, $\chi_{\rho}^{(1)}$ is associated with the change of the eigenvectors and $\chi_{\rho}^{(2)}$ the eigenvalues \cite{NoteOnNotation}. 
Note that the OFS may alternatively be expressed in terms of a dynamical two-point autocorrelation function of the perturbation \cite{ProsenLoshmidtDynamics}.

From the explicit form of $\chi_{\rho}^{(1,2)}(\lambda)$, one may identify how the OFS reflects the chaotic behavior of the system. 
In particular, we will be interested in the long-time behavior of the OFS.
Observe that for $\chi_{\rho}^{(1)}(\lambda)$, at large times $t$ the function $G_{t}(x)$ acts as a filter selecting  the contributions due to neighboring levels in the sum (\ref{eq:Chi1}). In fact, since $\genfrac{}{}{0pt}{}{\lim}{t\to\infty}G_{t}(x)=\delta(x)$,  the largest contributions to $\chi_{\rho}^{(1)}$ come from small gaps $s_n=E_{n+1}-E_{n}<1/t$. Therefore, for large $t$ $\chi_{\rho}^{(1)}(\lambda)$ becomes highly sensitive to the specific level spacing distribution that characterizes the model for a given $\lambda$. As for the second term of the OFS, it displays a quadratic dependence on time and may be written as $\chi_{\rho}^{(2)}(\lambda)= t^2 \Delta_\rho^2 V_{d}$, where 
$\Delta_{\rho}^2 V_{d} = \average{{V_{d}}^2}_{\rho} - \average{V_{d}}_{\rho}^2$ is the variance of the diagonal part $V_d$ of the perturbation $V$ in the energy eigenbasis. 

We remark here on the domain of validity of the OFS approximation for the fidelity.  In order for the expansion (\ref{eq:OFS}) to be valid, it is necessary that 
$\mathcal{F}_{\rho} \approx 1$. Since the rate of growth of the OFS is at most quadratic in time, it is thus necessary that $(t \delta \lambda)^2 \ll 1$. Provided 
this holds, it is also desired that the factor $G_t(E_n-E_m)$ in $\chi_{\rho}^{(1)}$ have a narrow-enough peak that it samples primarily the nearest-neighbor level 
spacings. If this is satisfied, $\chi_{\rho}^{(1)}$ is expected to be sensitive to the level spacing statistics. Given the set of energy levels in the support of $\rho$ and defining some typical level spacing 
$\Delta_{\textrm{typ}} = \langle E_{n+1}-E_{n} \rangle_{\textrm{typ}}$, 
the time scale in which $G_t(x)$ samples primarily the nearest-neighbor level spacings is given by $1/t \approx \Delta_{\textrm{typ}}$.

In our analysis we consider two kinds of initial states, restricting ourselves to a single parity sector of the model 
in order to allow comparison with the universal level spacing statistics.
On one hand we will use a maximally mixed state $\rho_D=\mathbb{I}/D$, defined as a truncated version of the exact state $\rho$ whose range will be the $D$-dimensional space of numerically well-converged states in the even sector.
On the other hand we consider $\rho$ to be a (truncated) Gibbs thermal state over the even sector, $\rho_{\lambda}=e^{-\beta H_{\lambda}}/Z_{\lambda}$, where $Z_{\lambda}=\textrm{Tr}(e^{-\beta H_{\lambda}})$. This will allow us to establish the extent to which the introduction of temperature in the system modifies the behavior of the OFS.

\section{\label{sec:Model}The model}

The hydrogen atom in a uniform external magnetic field is one of the
simplest time-independent systems exhibiting quantum chaos, and has
been studied in detail analytically, numerically, and experimentally
\cite{LesHouches, DG, DGJPhysB}. The Hamiltonian of a non-relativistic hydrogen
atom in a uniform cylindrically-symmetric external magnetic field
has the form \begin{equation}
H=\frac{p^{2}}{2}-\frac{1}{r}+\frac{\gamma}{2}L_{z}+\frac{\gamma^{2}}{8}(x^{2}+y^{2}),\label{eq:HOrig}\end{equation}
 where $\gamma$ is a dimensionless parameterization of the magnetic
field strength and $L_{z}$ is the component of angular momentum along
the magnetic field axis. This component of angular momentum is conserved,
and we restrict ourselves to the $L_{z}=0$ sector. For weak magnetic
fields a non-zero angular momentum quantum number simply results in
a uniform shift of the energy levels within a given sector, the Zeeman
effect.

The application of a magnetic field to the classical hydrogen atom
gives rise to the possibility of richly complex electron orbits. Two
principal orbital modes of the electron can be seen: a ``rotational''
mode in the $z=0$ plane and a ``vibrational'' mode along the
$z$-axis \cite{LesHouches}. An effect of a sufficiently strong magnetic
field is to make trajectories near these modes unstable. For the classical
model, it can be seen from scaling arguments that the degree of regularity
or chaos is determined chiefly by the single parameter $\widetilde{\epsilon}=E\gamma^{-2/3}$
given by the energy $E$ and magnetic field strength $\gamma$, where
the limiting cases are $\widetilde{\epsilon}\to-\infty$ ($\widetilde{\epsilon}\to\infty$)
for the Coulomb (Landau) regimes \cite{LesHouches}. The degree of
chaos is dependent on the relative strengths of the Coulomb attraction
to the nucleus and the diamagnetic interaction of the electron with
the magnetic field. For $\widetilde{\epsilon}$ less and less negative,
the perturbation due to the magnetic field begins to dominate, and
the phase space becomes increasingly chaotic. Note that the analysis
here considers only the bound spectrum.

Returning to the quantum mechanical hydrogen atom, with a suitable
change to ``semiparabolic'' coordinates $\mu=\sqrt{r+z}$ and
$\nu=\sqrt{r-z}$, the Schr\"odinger equation for the hydrogen atom
of energy $E$ may be written

\begin{equation}
\big[D_{\mu}+D_{\nu}-E(\mu^{2}+\nu^{2})+\frac{\gamma^{2}}{8} \mu^{2}\nu^{2}(\mu^{2} + \nu^{2})-2\big]\ket{\Psi}=0,\label{eq:HSemiparabolic}\end{equation}

where $D_{\mu}\equiv-\frac{1}{2}\big(\frac{\partial^{2}}{\partial\mu^{2}}+\frac{1}{\mu}\frac{\partial}{\partial\mu}\big)$, and similarly for $D_{\nu}$.

If one now performs another coordinate transformation to {}``dilated''
semiparabolic coordinates \cite{DG}, $u=(-2E)^{1/4}\mu$ and $v=(-2E)^{1/4}\nu$,
the Schr\"odinger equation takes the form

\[
\big[D_{u}+D_{v}+\frac{u^{2}+v^{2}}{2}+\frac{\gamma^{2}}{8(-2E)^{2}} u^{2}v^{2}(u^{2} + v^{2})-\frac{2}{\sqrt{-2E}}\big]\ket{\Psi}=0.\]

Here it can be seen that the hydrogen atom is equivalent to a pair
of coupled two-dimensional harmonic oscillators, each having identical angular
momenta which are in this case zero. If we now define $\lambda=\gamma^{2}/(-2E)^{2}$
and $\epsilon=(-2E)^{-1/2}$, the final form for the Schr\"odinger equation
is

\begin{equation}
\Big[\frac{D_{u}+D_{v}}{2}+\frac{u^{2}+v^{2}}{4}+\frac{\lambda}{16} u^{2} v^{2}(u^{2} + v^{2})\Big]\ket{\Psi}=\epsilon\ket{\Psi}.\label{eq:OscillatorHamiltonian}\end{equation}

The problem we aim to solve in this work is to find the eigenvalues
$\epsilon$ corresponding to the choice of coupling parameter $\lambda$.
This is called the \emph{oscillator} problem. Notice that our choice
of $\lambda$ fixes the ratio $\gamma/E$, so the oscillator problem
essentially corresponds to finding the hydrogen atom energies $E$
intersecting a curve of constant $\gamma/E$ in the $\gamma$-$E$
plane. Alternatively, one may solve the hydrogen problem for the energies
$E$ for a given magnetic field $\gamma$. This is a \emph{generalized}
eigenvalue problem of the form $\hat{A}\ket{\Psi}=\alpha\hat{B}\ket{\Psi}$
(Eq. \ref{eq:HSemiparabolic}). In Ref. \cite{DG} both approaches
have been taken, with similar results for the level spacing statistics.

\subsection{\label{sec:Algebraic-representation}Algebraic representation and
numerical procedure}

The problem of finding an efficient solution to the eigenvalue problem for the hydrogen atom and to the equivalent nonlinearly coupled harmonic oscillators has been discussed in several papers. In particular, Delande and Gay were able to solve the problem elegantly by giving a dynamical group approach representation of the model \cite{DGJPhysB}. In the following we briefly review their results. One begins by defining the operators

\begin{eqnarray}
S_{\genfrac{}{}{0pt}{}{x}{z}} & = & \frac{1}{4}\big[\pm\big(\frac{\partial^{2}}{\partial u^{2}}+\frac{1}{u}\frac{\partial}{\partial u}\big)+u^{2}\big]\\
S_{y} & = & \frac{i}{2}\big[1+u\frac{\partial}{\partial u}\big]\nonumber \end{eqnarray}
that generate the so(2,1) algebra and fulfill the commutation relations \cite{DGJPhysB} \begin{eqnarray}
\left[S_{x},S_{y}\right] & = & -iS_{z}\\
\left[S_{y},S_{z}\right] & = & iS_{x}\\
\left[S_{z},S_{x}\right] & = & iS_{y}.\end{eqnarray}
In terms of these operators, the Schr\"odinger equation (\ref{eq:OscillatorHamiltonian})
can be written as
\[
\big[S_{z}+T_{z}+\frac{\lambda}{2}V-\epsilon\big]\ket{\Psi}=0,\]
where 
\[
V=(S_{x}+S_{z})^{2}(T_{x}+T_{z})+(S_{x}+S_{z})(T_{x}+T_{z})^{2}
\]
and $T_{x,y,z}$ has the same form as $S_{x,y,z}$, except with
$u$ replaced by $v$. Note that $[T_{\alpha},S_{\beta}]=0$  $\forall$ $\alpha,\beta\in \{x,y,z\}$.
With this representation, the natural basis to use is the tensor product of the eigenbases of
$S_{z}$ for each oscillator. That is, the set of states $\{\ket n\otimes\ket m\}_{n,m\in\mathbb{N}}$ where
\begin{eqnarray}
S_{z}\ket n & = & \big(n+\frac{1}{2}\big)\ket n\label{eq:Basis}\\
S^{+}\ket n & = & (n+1)\ket{n+1}\nonumber \\
S^{-}\ket n & = & n\ket{n-1}\nonumber \end{eqnarray}
and $S^{\pm}\equiv S_{x}\pm iS_{y}$. In this representation, $S_{\alpha} \to S_{\alpha} \otimes \mathbb{I}$ and $T_{\alpha} \to \mathbb{I} \otimes S_{\alpha}$.
The Hamiltonian for the oscillator problem may then be written
\begin{equation}
H_{\lambda}=S_{z}\otimes1+1\otimes S_{z}+\lambda V,\label{eq:Hfinal}\end{equation}
where
\begin{equation}
V=\frac{1}{2}\big[\Sigma^{2}\otimes\Sigma+\Sigma\otimes\Sigma^{2}\big]\label{eq:Perturbation}\end{equation}
and $\Sigma\equiv S_{x}+S_{z}=S_{z}+\frac{1}{2}(S^{+}+S^{-})$. Notice
that the Hamiltonian is symmetric under interchange of the subsystems.
Namely, defining the parity (or swap) operator $P$ acting in the oscillator occupation basis
as $P\ket{n} \otimes \ket{m} = \ket{m} \otimes \ket{n}$, we have $[H,P]=0$.

With this in mind, we now define an orthonormal basis
\[
\ket{e_{n,m}}=\left\{ \begin{array}{ll}
\frac{1}{\sqrt{2}}\big(\ket n\otimes\ket m+\ket m\otimes\ket n\big) & (n\neq m)\\
\ket n\otimes\ket n & (n=m)\end{array}\right.\]
 \[
\ket{d_{n,m}}=\frac{1}{\sqrt{2}}\big(\ket n\otimes\ket m-\ket m\otimes\ket n\big)\quad(n\neq m),\]

where $\big\{\ket{e_{n,m}}\big\}$ $\big(\big\{\ket{d_{n,m}}\big\}\big)$
span the even (odd) parity subspaces. Since the parity operator $P$ and Hamiltonian $H$ commute, the
Hamiltonian can be block diagonalized over the even and odd parity subspaces. From now on we restrict ourselves
to states $\rho$ which have support only over the even subspace, in order to better compare the fidelity analysis with the level spacing statistics. Indeed, in general for studies of quantum chaos it is useful to identify the symmetries of the problem and focus on individual symmetry subsectors.  In the chaotic regime, for example, if the energy levels from several symmetry subspaces are included the level spacing statistics may take a non-universal form arising from the superposition of respectively universally-distributed subsets of levels.

While in general the system to be analyzed has to be represented in an infinite-dimensional Hilbert space $\mathcal{H}$, in order to handle the problem numerically one must truncate the original Hamiltonian into one which acts on a finite-dimensional space $\mathcal{H}_{K}$ and appropriately represents
the low-energy physics of the exact model. The above described algebraic representation of the pair of two-dimensional harmonic oscillators easily allows for such a truncation. Here, the Hilbert space truncation is implemented as an upper bound on the total allowed oscillator occupation, requiring $n+m\leq K$, where
$K$ is our truncation parameter. The following is the truncated
Hilbert space dimension for the even sector in terms of $K$:
\[
\textrm{Dim}(\mathcal{H}_{K}^{\textrm{even}})=\begin{cases}
\big(\frac{K}{2}+1\big)^{2}, & K\;\textrm{even}\\
\big(\frac{K+1}{2}\big)\big(\frac{K+1}{2}+1\big), & K\;\textrm{odd}.\end{cases}\]

However, the eigenenergies of the truncated Hamiltonian poorly approximate the exact values near the top of the truncated spectrum. We address the inaccuracies introduced
by the truncation by extending the truncation and measuring the corresponding
changes in the energy spectrum, so that for a given truncation size
we may determine the set of levels which are sufficiently well-converged
to the true values for our purposes. We find that as the coupling
grows the necessary truncation dimension also must increase
to provide the same degree of convergence. This effect is due to shielding
from the influence of the magnetic field due to strong coulomb attraction
for lower energy levels. For weaker magnetic fields a larger number
of eigenvalues will be well-converged, but to be consistent we include
only those levels which are well-converged over the entire coupling
interval of interest.

Recall that the parametrization of the coupling between the oscillators, $\lambda$, is related to the hydrogen atom problem through $\lambda=\gamma^{2}/(-2E)^{2}$, where $\gamma$ and $E$ were the external magnetic field strength and energy of the hydrogen atom, respectively. In this work, since we are studying the oscillator problem we only vary the oscillator coupling $\lambda$ and do not independently adjust the external magnetic field parametrizing the hydrogen problem.
As previously mentioned, the quantity $\eta=\lambda\epsilon^{2}$ was found to characterize the degree to which the phase space of the
classical model is chaotic \cite{DG}. In particular, for $\eta \lesssim 1$ the phase space is almost entirely regular while for $\eta \approx 60$ the phase 
space becomes entirely chaotic \cite{DG}. It is known for this model that the character of the level spacing statistics is reflected in the proportion of the 
classical phase space which is chaotic. 
As we will observe now, where $\eta$ is small the level statistics is Poissonian, $P(s)=e^{-s}$,
while for larger $\eta$ the level statistics transforms into a Wigner-Dyson form,
$P(s)=\frac{\pi}{2}se^{-\frac{\pi}{4}s^{2}}$.

\section{\label{sec:Results}Results}

\subsection{\label{sec:Level-spacing-statistics}Level spacing statistics}
\begin{figure}[htp]
\includegraphics[scale=0.36]{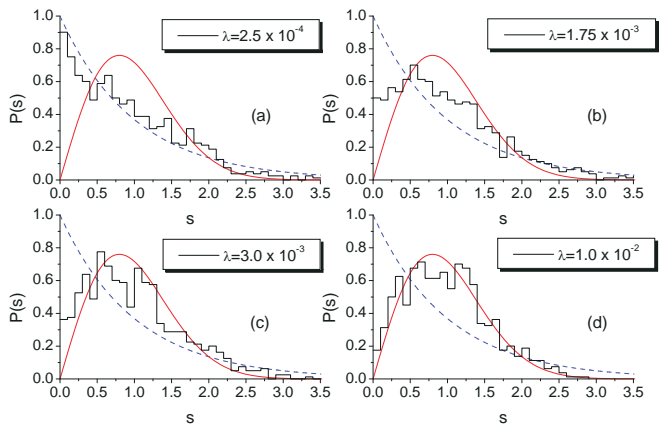} \caption{(Color online) Level spacing distributions for: (a) $\lambda=2.5 \textrm{ x } 10^{-4}$,
approximately the location of the local minimum of $\chi^{(1)}(\lambda)$,
(b) $\lambda=1.75 \textrm{ x } 10^{-3}$, approximately the peak of $\chi^{(1)}(\lambda)$,
(c) $\lambda=3.0 \textrm{ x } 10^{-3}$, in the tail of the peak, and (d) $\lambda=1.0 \textrm{ x } 10^{-2}$,
far into the quantum chaotic regime. Here all $800$ well-converged
levels are included in the statistics. The Poisson statistics are
given by $P(s)=e^{-s}$ (dashed curve) and the Wigner-Dyson statistics are $P(s)=\frac{\pi}{2}se^{-\frac{\pi}{4}s^{2}}$ (solid curve).}
\label{fig:SpecStat}
\end{figure}

With these well-converged levels in hand, we now compute the level spacing
statistics and compare them with the results of Delande and Gay \cite{DG}.
Gradually increasing the coupling from the
perturbative regime (Fig. \ref{fig:SpecStat}a) through the so-called
\emph{n-mixing} regime, where levels from neighboring sectors begin to cross
(Fig. \ref{fig:SpecStat}b) until the quantum chaotic regime (Fig.
\ref{fig:SpecStat}c and \ref{fig:SpecStat}d), the
level spacing statistics smoothly evolve from Poisson-like to Wigner-Dyson-like.
Note that whereas the level statistics presented in \cite{DG} are
the superimposed statistics for several coupling strengths, we have
evaluated the level spacings for individual couplings.
Our statistics include 800 well-converged eigenvalues, with a truncation
of $K=120$.

\begin{figure}[hbp] \includegraphics[scale=0.45]{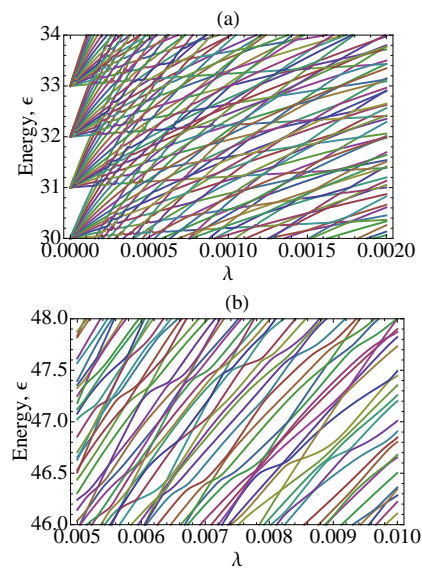}\caption{\label{fig:LevelTracing}(Color online) Eigenenergies vs. coupling strengths for $K=120$ for (a) couplings $0<\lambda<2 \textrm{ x } 10^{-3}$ and energies $30<E<34$, such that $0<\eta<2.3$, and (b) couplings $5 \textrm{ x } 10^{-3}<\lambda<10^{-2}$, such that $10<\eta<23$.} \end{figure}

In Fig. \ref{fig:LevelTracing} the eigenenergies are plotted as a
function of coupling for two representative intervals
of couplings and energies. The changing character of the energy spectrum
as a function of the coupling strength is evident. For small couplings,
degeneracies are lifted and levels from neighboring sectors begin
to cross as the n-mixing regime is entered. As the
coupling strength increases further still, the regular n-mixing regime
evolves into quantum chaos as level avoidance begins to dominate.

\subsection{The $\chi^{(1)}$ term}

Having reviewed the behavior of the level spacings, let us now examine the operator fidelity susceptibility.
We start by analyzing $\chi^{(1)}$, first discussing the cut-off approximation:
\begin{equation}
\chi^{(1)}(D_c)=\frac{2 \pi t}{Z(D_c,\beta)}\sum_{n=1}^{D_c}\exp(-\beta E_{n})C_{D_c}(n),
\label{eq:Chi1Cutoff}
\end{equation}
where $C_{D_c}(n)\equiv\sum_{m=1,m\neq n}^{D_c}|\bra nV\ket m|^{2}G_{t}(E_{n}-E_{m})$,
$\beta=1/T$, and $Z(D_c,\beta)\equiv\sum_{n=1}^{D_c}\exp(-\beta E_{n}).$
Here $\{\ket n\}$ are the $(\frac{K}{2}+1)^{2}$-dimensional or $(\frac{K+1}{2})(\frac{K+1}{2}+1$)-dimensional
(for $K$ even, odd respectively) energy eigenvectors of the even parity block of the Hamiltonian
$H_{K}$. The number of states included in the sum is called the cut-off dimension, $D_c$. Note that if the odd sector were included as well, since the perturbation $V$ preserves parity the sum in (\ref{eq:Chi1Cutoff}) would be simply the linear combination of $\chi^{(1)}$ computed over each parity sector. The dependence of all factors on both the coupling $\lambda$
and truncation dimension $K$ has been left implicit.
In our calculations we include only the states corresponding to well-converged
eigenvalues, namely levels which vary little under additional growth
of the truncation dimension $K$. Taking $K=120$, for example, we retain $D_c=800$
out of the $3721$ total levels in the even parity sector for the strongest coupling
considered, $\lambda=10^{-2}$. Again, though the number of well-converged
levels is maximal for smaller couplings, for consistency in our calculations we include the worst-case set of levels for the entire region of interest.

For finite temperature, contributions to the OFS from higher energies
are exponentially suppressed due to the coefficient $\rho_{n}=\exp(-\beta E_{n})$
in the outer sum of Eq. (\ref{eq:Chi1Cutoff}). This serves to give more weight
to the low-energy part of the spectrum for low temperatures, while
for high-enough temperatures all levels are included.
\begin{figure}
\includegraphics[scale=0.3]{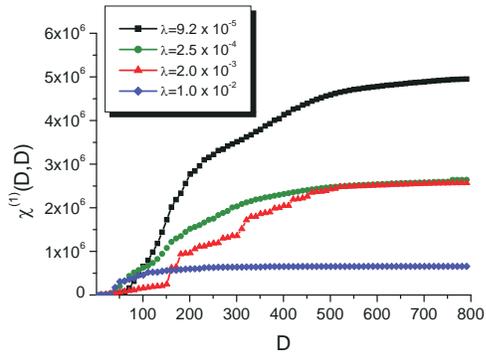}
\caption{\label{fig:Chi1_Convergence}(Color online) Partial sum $\chi^{(1)}(D,D)$
as a function of D, for temperature $T=4.5$ and $\lambda=9.2 \textrm{ x } 10^{-5}, 2.5 \textrm{ x } 10^{-4}, 2.0 \textrm{ x } 10^{-3}, \textrm{ and } 1.0 \textrm{ x } 10^{-2}$.}
\end{figure}
In Fig. \ref{fig:Chi1_Convergence} we show the partial sum $\chi^{(1)}(D_c)$
as a function of $D_c$ for up to $800$ well-converged energy levels,
for four different couplings $\lambda$. For example, we can see that a temperature
of about $T=4.5$ is sufficient to provide a well-converged result
for $\chi^{(1)}$, with the result being better-converged for
larger coupling. This is due to an expansion of the energy spectrum towards larger energies
as the perturbation strength increases. From now on, we define $\chi^{(1)}$ as the partial
sum (\ref{eq:Chi1Cutoff}) with $D_c=800$.

\begin{figure}[hbp]
\includegraphics[scale=0.3]{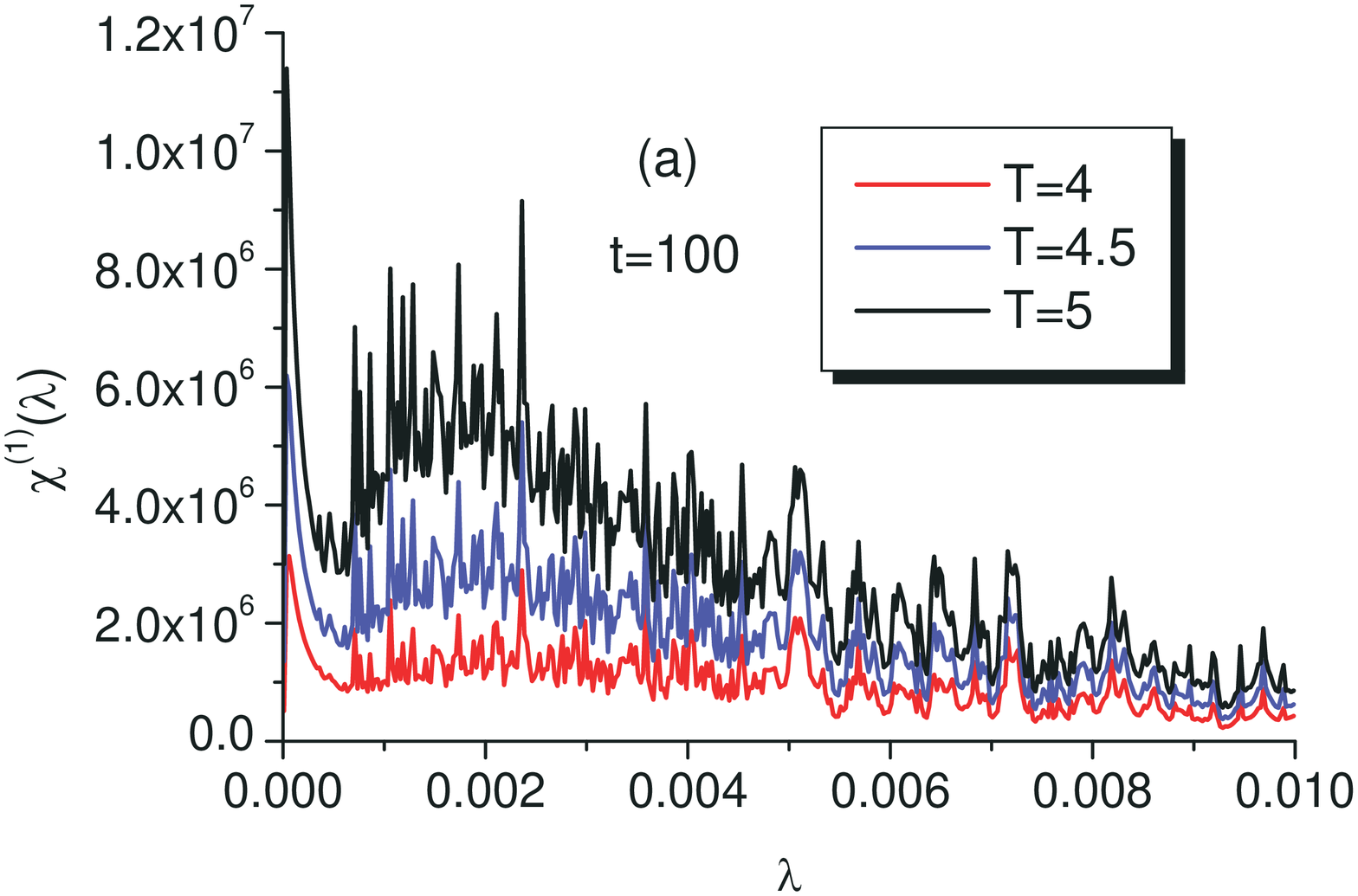} \vskip0mm
\includegraphics[scale=0.3]{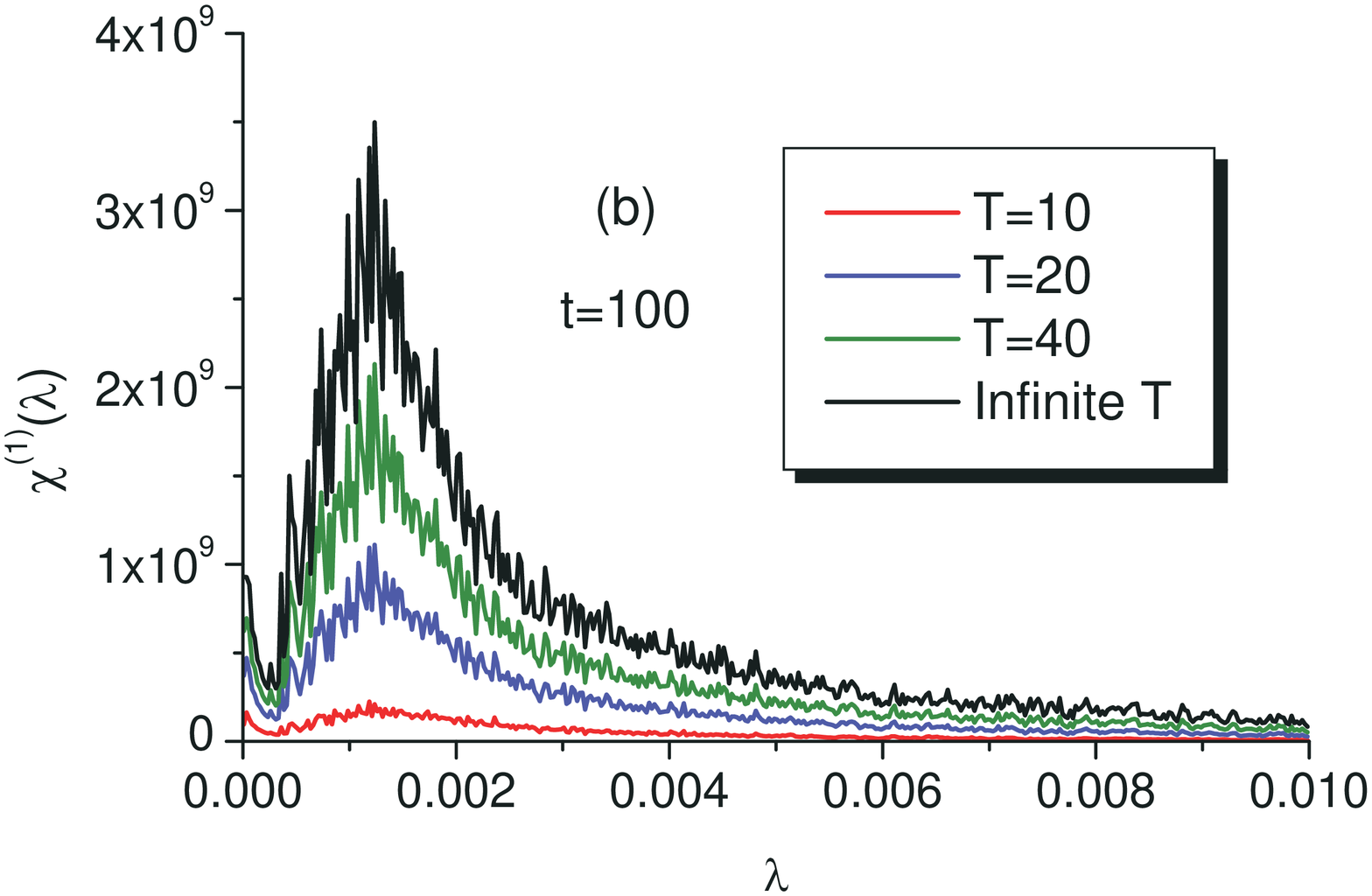}
\caption{\label{fig:Chi1_t100}(Color online) The first term of the OFS, $\chi^{(1)}(\lambda)$, with $K=120$, 
$D_{c}=800$, and $t =100$ for (a) small temperatures and (b) higher temperatures.}
\end{figure}

Let us now look at the parameter dependence of the quantity $\chi^{(1)}(\lambda)$
for finite temperature, plotted in Fig. \ref{fig:Chi1_t100}. Recall
from Ref. \cite{LSWZ} that $\chi^{(1)}(\lambda)$ characterizes the
variation of the eigenstates, and as shown in Eq. (\ref{eq:Chi1}) is
a function of all level spacings or gaps. The plot is jagged
due to the strong contributions of individual level crossings or near
misses. However, one may identify two inflection points. For small
couplings there is a local minimum of $\chi^{(1)}$, while for progressively
larger couplings a global maximum appears. Past the global maximum,
$\chi^{(1)}$ decreases as level avoidance begins to appear. From
then on, as the couplings grow further $\chi^{(1)}$ continues to
decrease.

Our explanation is the following. For vanishing coupling the nearly
degenerate subsectors of the spectrum give a large contribution to
$\chi^{(1)}$ due to small arguments to $G_{t}(E_{n}-E_{m})$ and
significant couplings $|\bra nV\ket m|$ between neighboring levels.
However, a slight increase of the coupling lifts those degeneracies,
taking the system to the regime where perturbation theory applies
since level crossings are rare (see Fig. \ref{fig:LevelTracing}a),
and consequently $\chi^{(1)}$ is reduced. Continuing to increase
the coupling, the energy levels between subsectors now begin to cross
as one enters the so-called \emph{n-mixing} regime. With sufficiently
large coupling the system finally enters the quantum chaotic regime,
where couplings between neighboring levels now lead to avoided level
crossings, infrequent level crossings, and a consequent decrease in
$\chi^{(1)}$. Note that the perturbative regime, for example, persists
for a wider range of couplings for low energies than for higher energies,
so the system is never entirely in one regime or another. This can
be seen by comparing the energies and couplings with the classical
chaos parameter $\eta=\lambda\epsilon^{2}$. For the hydrogen atom,
higher energies result in a greater susceptibility of the electron
to the external magnetic field over the Coulomb potential.

\begin{figure}[hbp]
\includegraphics[scale=0.3]{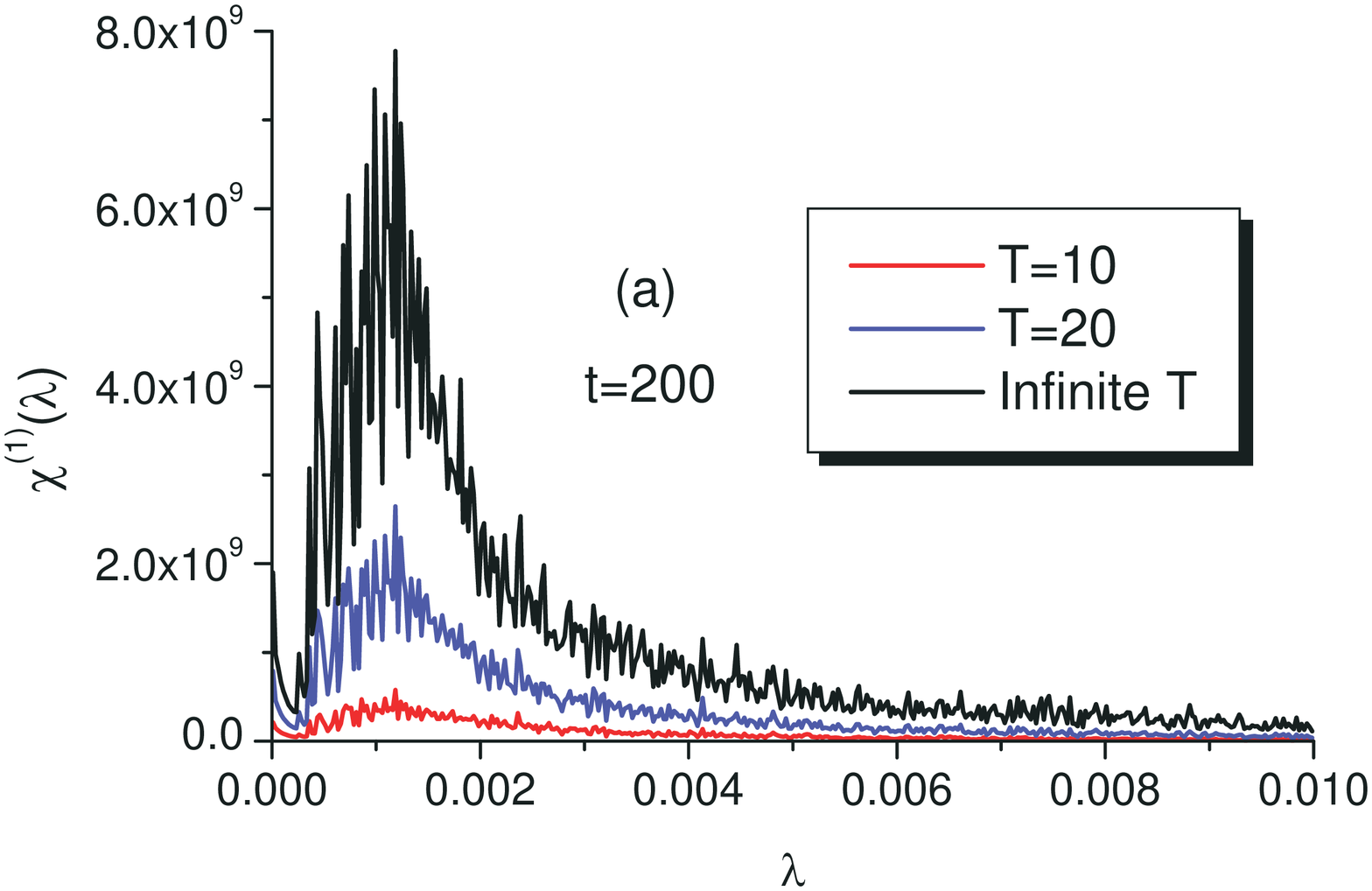} \vskip0mm
\includegraphics[scale=0.3]{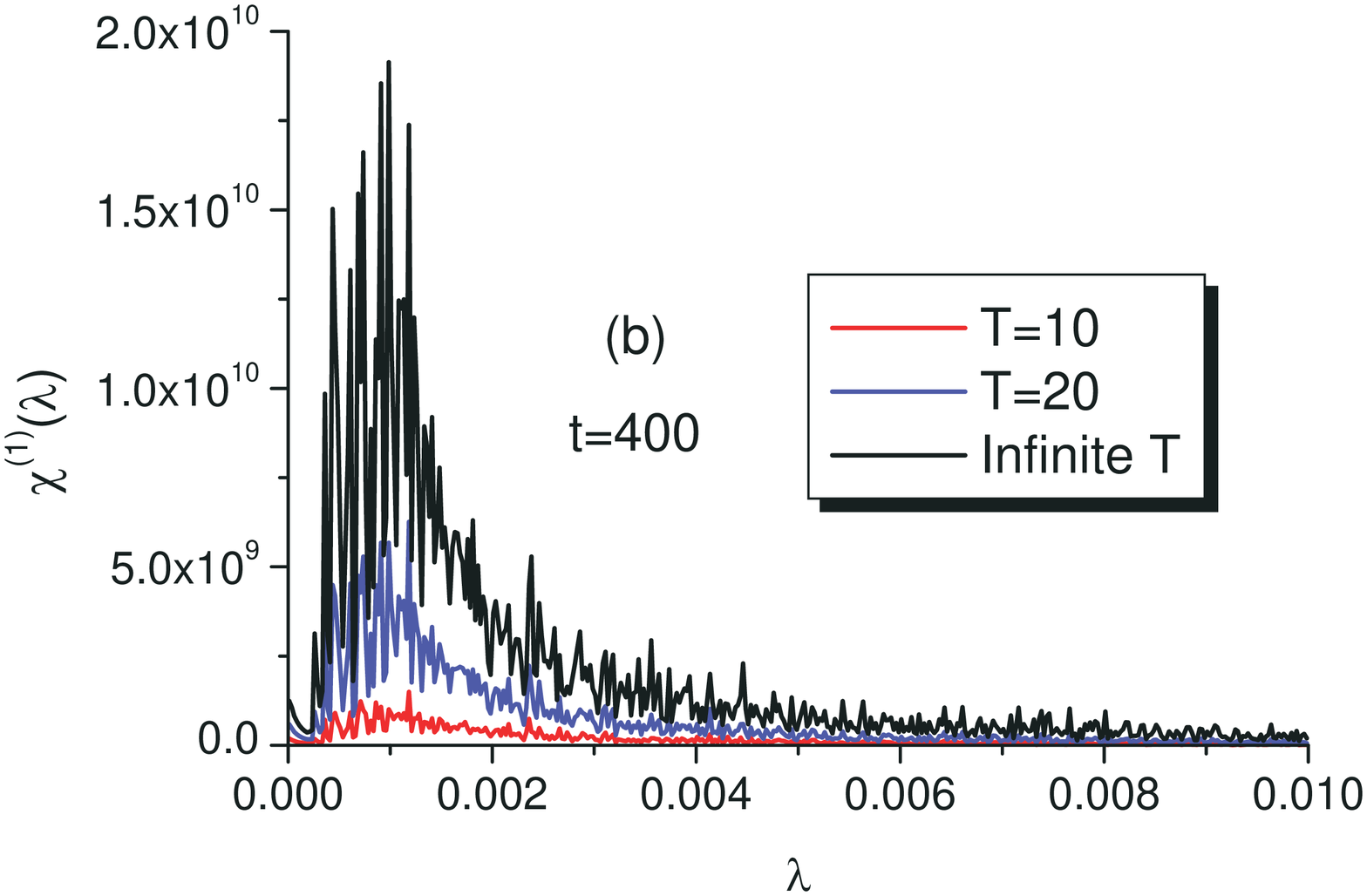}
\caption{\label{fig:Chi1_t200400}(Color online) The first term of the OFS, $\chi^{(1)}(\lambda)$, with $K=120$, 
$D_{c}=800$, and various high temperatures at (a) $t=200$ (b) $t=400$.}
\end{figure}

In addition to taking a finite-temperature Gibbs state $\rho=e^{-\beta H}/Z$,
another natural choice for the density matrix $\rho$ is to consider
the limit $T\to\infty$. In this case our density matrix becomes proportional
to a projector $P$ over the subspace spanned by those eigenstates
having well-converged eigenvalues. Acting on this well-converged subspace,
the state will have the form $\rho_{\infty}=\mathbb{I}/D_{c}$. The
result for $\chi_{\infty}^{(1)}$ is plotted as a function of $\lambda$
in Figs. \ref{fig:Chi1_t100} and \ref{fig:Chi1_t200400}. With this form of the state we see that
the contrast between the various regimes is enhanced. Moreover, this state uniformly weights the contributions of many more levels than the low-temperature case, suggesting that variation of $\chi_{\infty}^{(1)}$
may more closely mirror the level spacing statistics. It
turns out that increasing the temperature strictly increases the magnitude
of $\chi^{(1)}$, so that the ``infinite''-temperature case of $\rho=\mathbb{I}/D_{c}$
has the largest value. Moreover, the location of the initial local
minimum for small couplings shifts toward smaller values as temperature
is increased. This is due to the sampling of higher energy levels,
for which smaller couplings are required to traverse the various regimes.

We can thus conclude that, as has already been found for the Dicke model in \cite{GZ}, the $\chi^{(1)}$ part of the OFS correctly incorporates information about the level spacing statistics for the case of the nonlinearly coupled oscillators. As shown in Fig. \ref{fig:Chi1_t200400}, modifying the time $t$ does not qualitatively change the plot of
$\chi_{\rho}^{(1)}$. Noting the definition of $G_{t}(x)$ following
equation (\ref{eq:Chi1}), where $\lim_{t\to\infty}G_{t}(x)=\delta(x)$,
a larger time enhances the contributions to the sum due to level crossings or quasi-
level crossings. The growth of $\chi^{(1)}$ for larger times thus reflects its sensitivity to the presence (absence) of level crossings, and hence the transition from regularity to chaos.

\subsection{The $\chi^{(2)}$ term}

\begin{figure}
\includegraphics[scale=0.32]{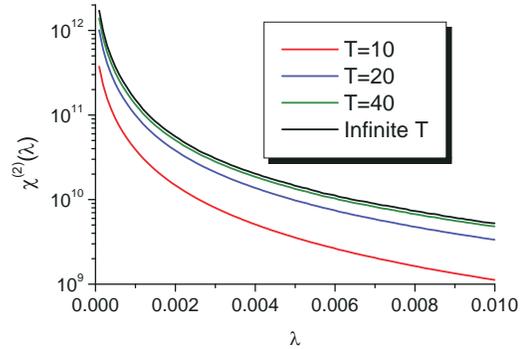}
\caption{\label{fig:Chi2}(Color online) The second term of the OFS, $\chi^{(2)}(\lambda)$, taking $K=120$, 
$D_{c}=800$, and $t=100$}
\end{figure}

So far we have treated the part of the OFS, $\chi^{(1)}(\lambda)$,
which reflects the variation of the eigenstates and depends on the
level spacings. The other term, $\chi^{(2)}(\lambda)$, is proportional
to the variance of the diagonal elements of the perturbation $V=\partial_{\lambda}H$
with respect to $\rho$ in the energy basis. In Fig. \ref{fig:Chi2} one sees that $\chi^{(2)}$ rapidly decreases with coupling strength. The values of $\chi^{(1)}$ and $\chi^{(2)}$ in both the regular and chaotic regimes differ by about two orders of magnitude, and this is not qualitatively affected by temperature change.
This consideration allows us to conclude that for this model the dominant contribution to the overall OFS is given by the 
fluctuations of the diagonal part of the perturbation $\hat{V}$.
However, though the contribution from $\chi^{(1)}$ in this case is relatively small, it possesses a richer functional dependence on the coupling. In particular, 
while $\chi^{(2)}$ decreases monotonically in the studied parameter interval, $\chi^{(1)}$ reflects the transition through the three described regimes.

In the special case where the diagonal part of the perturbation in the energy basis, $V_d$, is zero, this second term will clearly 
vanish \cite{ProsenLoshmidtDynamics}. 
Thus, for an experiment or theoretical treatment in which it is desired that $\chi_{\rho}^{(1)}$ be isolated, choosing a perturbation of this form will eliminate $\chi_{\rho}^{(2)}$.

Our results show that the behavior of the total OFS as a function of the parameter that drives the transition from the regular regime to a chaotic one encodes the 
already-mentioned resilience of the quantum evolutions to small non-random perturbations even for specific systems. Indeed, for our model the statistical distinguishability between neighboring evolutions decreases with $\lambda$, showing that the resilience of the system to perturbations dramatically increases when it becomes chaotic.

\section{\label{sec:Conclusion}Conclusion}

In this work we have used an operator-geometric quantity, the operator
fidelity susceptibility (OFS), to study the transition from regularity
to quantum chaos for a pair of coupled two-dimensional harmonic oscillators.
This model is equivalent to a hydrogen atom in a uniform external
magnetic field, one of the prototypical systems for which both the regimes of classical and quantum chaos have been well-documented theoretically and experimentally.
We have seen that, by computing the state-dependent part of the OFS, denoted $\chi^{(1)}(\lambda)$, as a function of the coupling strength
$\lambda$ one may distinguish three regimes in parameter space: perturbative, n-mixing, and quantum chaotic. A local minimum for small coupling strength corresponds to the boundary between the perturbative and n-mixing regimes, while the global maximum
coincides with the occurrence of level crossings typical of a regular regime. As the level spacing statistics transform from 
Poisson (maximum likelihood of level crossing) to Wigner-Dyson (vanishing probability of level crossings), $\chi^{(1)}(\lambda)$ correspondingly decreases.
We may therefore conclude that $\chi^{(1)}(\lambda)$ both incorporates the information relative to the statistics of the spacings between neighboring energy levels and the information-theoretic notion of distinguishability between quantum evolutions. 
Our analysis shows that both the distinct elements of the OFS and the OFS as a whole are therefore effective tools for discriminating the different characters, i.e. regular vs. chaotic, of the various regimes of a system's parameter space.

We would like to remark that while the OFS approach may not be as efficient computationally as other more direct techniques, i.e. 
the level spacing statistics, the main goal in this work is to see whether the notion of 
resilience to perturbation, quantified by the OFS, is a useful one in the context of this well-known model.  
In the future, it would be interesting to experiment with this approach on models where the connection between 
classical chaos and energy level statistics breaks down, such as in the odd-parity sector of the lithium atom \cite{CourtneyKleppner}. 
Indeed, does the ability of $\chi_{\rho}^{(1)}$ to detect the transition from regularity to chaoticity solely depend upon a corresponding change in the level spacing statistics, or can it serve as a measure for chaoticity in systems without this property?

\begin{acknowledgements}
We are grateful for the facilities and assistance of USC's High Performance Computing and Communications center, where all of the numerical calculations were performed.
This work has been supported by NSF grants: PHY-803304, DMR-0804914.
\end{acknowledgements}


\begin{thebibliography}{15}
\bibitem{NoteOnNotation} Note that the notation for labeling $\chi^{(1,2)}$ here is the reverse of that of Ref. \cite{LSWZ}, but is consistent with that of Ref. \cite{GZ}.
\end{thebibliography}
\end{document}